\documentclass[
  aps,
  prd,
  onecolumn,
  superscriptaddress,
  nofootinbib,
  longbibliography,
  floatfix,
  amsmath,
  amssymb,
  10pt
]{revtex4-2}

\usepackage[T1]{fontenc}
\usepackage[utf8]{inputenc}
\usepackage{newtxtext,newtxmath}
\usepackage{microtype}
\usepackage[dvipsnames]{xcolor}
\usepackage{amsmath,amssymb,amsfonts}
\usepackage{mathtools}
\usepackage{bm}
\usepackage{graphicx}
\usepackage{booktabs}
\usepackage{array}
\usepackage{orcidlink}
\usepackage{hyperref}
\usepackage{bookmark}
\usepackage{flafter}
\usepackage{placeins}
\usepackage{needspace}
\allowdisplaybreaks
\definecolor{darkblue}{rgb}{0,0,0.5}
\hypersetup{
  pdfstartview={FitH},
  pdftitle={Dark matter environments and safeguards for spacetime inference from horizon scale interferometry},
  pdfauthor={Mohsen Fathi},
  pdfsubject={Dark matter environments, black hole spacetime inference, closure statistics, model adequacy, and numerical convergence},
  pdfkeywords={dark matter, black holes, Event Horizon Telescope, M87, spacetime tests, tidal charge, closure quantities, model misspecification, numerical convergence},
  colorlinks=true,
  breaklinks=true,
  linkcolor=darkblue,
  citecolor=darkblue,
  urlcolor=darkblue
}

\graphicspath{{./}{figures/}}

\newcommand{\dd}{\mathrm{d}}
\newcommand{\DM}{\mathrm{DM}}

\newcommand{\Sig}{\Sigma}
\newcommand{\Rpot}{\mathcal{R}}
\newcommand{\Qcar}{\mathcal{Q}}
\newcommand{\Cmat}{\mathbf{C}}
\newcommand{\Jmat}{\mathbf{J}}
\newcommand{\Pperp}{\mathbf{P}_{\perp}}
\newcommand{\muv}{\bm{\mu}}
\newcommand{\rvec}{\mathbf{r}}

\begin{document}

\title{Dark matter environments and safeguards for spacetime inference from horizon scale interferometry}

\author{Mohsen Fathi\orcidlink{0000-0002-1602-0722}}
\email{mohsen.fathi@ucentral.cl}
\affiliation{Centro de Investigaci\'{o}n en Ciencias del Espacio y F\'{i}sica Te\'{o}rica (CICEF), Universidad Central de Chile, La Serena 1710164, Chile}

\date{\today}

\begin{abstract}
Horizon scale interferometry can test a black hole spacetime only when the data, source model, and numerical response are reliable. We study this requirement with the public 2017 M87* closure data, a frozen semi analytic emission model, explicit dark matter controls, and a rotating tidal charge deformation. We normalize NFW and Einasto halos, an adiabatic spike, a capture suppressed spike, and a heated crest for M87*. Even the intentionally optimistic rendered case, $M_{\rm DM}(<10M)/M_{\rm BH}=7.33\times10^{-5}$, changes the normalized image and visibility by only about $2.65\times10^{-6}$ and $4.9\times10^{-7}$. We then build an independent closure phase and log closure amplitude likelihood with covariance and three fixed high/low band correlation cases. Synthetic Kerr tests recover the expected statistic, coverage, and false positive rate. The real data give $\chi^2/N=1.7583$, $1.7206$, and $1.6873$, above the global adequacy limit of $1.5$; the worst band gives $2.0216$. Removing the most influential scan still leaves $\chi^2/N=1.6472$. If tidal charge is allowed anyway, the residual projects strongly onto it, but the preferred direction changes sign between image resolutions. Direct libraries at $N=192$, 224, and 256 also fail the differential response convergence tests. At a smoothing width of $0.5M$, the Kerr image changes by about $0.45\%$ between $N=192$ and 224, while the tidal charge response changes by about $49\%$. We therefore report no posterior or bound. Spacetime inference should remain closed until the adopted data and covariance are validated, the undeformed source passes an absolute adequacy test, and the differential metric response converges independently of the image.
\end{abstract}

\maketitle

\section{Introduction}\label{sec:intro}

A horizon scale image is not made by the spacetime alone. It also depends on photon propagation, plasma emission and absorption, source variability, calibration, sparse Fourier coverage, imaging choices, and numerical settings. This was already clear in the first millimeter very long baseline interferometry detections of compact structure in Sgr A* and M87. It became even clearer with the Event Horizon Telescope (EHT) images, polarization maps, and multi epoch studies of M87* and Sgr A* \cite{Fish2011,Doeleman2012,Akiyama2015,EHT2019I,EHT2019II,EHT2019III,EHT2019IV,EHT2019V,EHT2019VI,EHT2021VII,EHT2021VIII,EHT2022I,EHT2022II,EHT2022III,EHT2022IV,EHT2022V,EHT2022VI,EHT2024PersistentI,EHT2025PersistentII}. A larger ground array and future space baselines will improve the angular resolution and time coverage, but they will also make model checks more important \cite{Doeleman2019ngEHT,Blackburn2019,Doeleman2023ngEHT,Johnson2023ngEHT}.

Interferometric inference also has its own limits. Closure phases and closure amplitudes reduce station based calibration errors, but the independent closure basis and its covariance must be treated consistently \cite{Jennison1958,CornwellWilkinson1981,PearsonReadhead1984,NarayanNityananda1986,Thompson2017,Blackburn2020}. Image reconstruction and direct model fitting can react differently to sparse coverage, regularization, likelihood choices, and extended emission \cite{Chael2016,Chael2018Imaging,EHT2019III,EHT2019IV,LockhartGralla2022,Georgiev2026}. Source variability can also change closure quantities and fitted orientation parameters from one scan or epoch to another \cite{Broderick2016,Satapathy2022,EHT2024PersistentI,EHT2025PersistentII,Chang2025}. A likelihood can therefore be statistically correct and still give a misleading physical result when the source model is not adequate.

The source model is just as important. Modern black hole image calculations use general relativistic magnetohydrodynamic flows, electron heating models, polarized radiative transfer, and several independent ray tracing and transfer codes \cite{FishboneMoncrief1976,DeVilliersHawley2003,Gammie2003,McKinneyGammie2004,Noble2006,Tchekhovskoy2011,Dexter2009,Dolence2009,Younsi2012,Vincent2011,Chan2013,Moscibrodzka2016,Bronzwaer2018,Chael2018,Porth2019,Gold2020,Bronzwaer2020,Wong2022}. These studies show that the ring diameter, brightness asymmetry, polarization, subring structure, and variability are not set by geometry alone \cite{EHT2019V,EHT2021VII,EHT2021VIII,Johnson2020,Himwich2020,Desire2025}. A compact semi analytic source is still useful for a controlled test, but it should first give an adequate description of the real data.

The geometry has also been studied in great detail. The Kerr photon region and shadow follow from separable null geodesics and have been examined through analytic lensing, critical curves, photon rings, and parameterized departures from Kerr \cite{Carter1968,Bardeen1973,Chandrasekhar1983,Falcke2000,Bozza2002,HiokiMaeda2009,Schneider1992,Petters2001,Grenzebach2014,Grenzebach2015,JohannsenPsaltis2010II,Psaltis2018,CunhaHerdeiro2018,GrallaLupsasca2020,GrallaMarrone2020,PerlickTsupko2022}. EHT measurements have encouraged direct tests of strong gravity and many studies of non Kerr metrics. However, a visible difference between two images is not by itself a measurement of a spacetime deformation \cite{Bambi2017,Psaltis2020,Vagnozzi2023,Younsi2021,Amarilla2012,Eiroa2018,Neves2020,Hou2021,Guo2026Hayward}. The important quantity is the response in data space after nuisance freedom is included. This small response must be checked more strictly than the image itself.

Dark matter is part of the environment of every supermassive black hole, but its gravity near the horizon depends strongly on the density profile and on the history of the system. Smooth NFW, Einasto, Burkert, Hernquist, and steeper cusp models give very different inward extrapolations \cite{Einasto1965,Hernquist1990,Navarro1996,Navarro1997,Moore1998,Burkert1995,Zhao1996,Bertone2005,Errani2023cNFW}. Adiabatic black hole growth can form a steep spike, while capture, annihilation, mergers, stellar scattering, and self interactions can weaken or reshape it \cite{GondoloSilk1999,Ullio2001,Merritt2002,Merritt2004,BertoneMerritt2005,GnedinPrimack2004,Merritt2007,Sadeghian2013,FieldsShapiroShelton2014,Lacroix2015,FerrerWill2017,AlvarezYu2021,ShapiroHeggie2022,Bertone2024}. Some models also predict dark matter effects on emission or spacetime observables, and several studies have examined shadows and horizon scale signatures in dark matter environments \cite{Jusufi2019,Xu2018,Haroon2019,Errehymy2026Einasto}. Therefore, a radial mass term should not be introduced as an unnamed environment. Its profile, normalization, inner cutoff, and physical meaning must be stated clearly.

The deformation used here is a braneworld tidal charge. It comes from the effective four dimensional equations of extra dimensional gravity and has the Kerr--Newman algebraic form, although it is not an electromagnetic charge \cite{RandallSundrum1999a,RandallSundrum1999b,Shiromizu2000,Dadhich2000,Kanti2004,Maartens2010,Aliev2005,AlievTalazan2009}. Its optical properties, photon regions, accretion signatures, and possible shadow constraints have been studied in several settings \cite{ScheeStuchlik2009,Amarilla2012,Eiroa2018,Neves2020,Zakharov2022}. It is a useful one parameter test, but an apparent preference for it may also absorb an error in the source model or in the numerical response.

This is a model validation problem. Posterior predictive checks, goodness of fit, boundary calibration, and numerical verification answer different questions \cite{Chernoff1954,White1982,GelmanMengStern1996,BayarriBerger2000,SelfLiang1987,Protassov2002,Andrae2010,KennedyOHagan2001,KleijnVanDerVaart2012,Celik2008,Roy2005,OberkampfRoy2010}. A successful synthetic test validates the statistical method only for the assumed mean model. It does not show that the source model is adequate for real observations. In the same way, a stable looking image does not show that the much smaller change produced by a metric deformation has converged.

\begin{figure}[t]
\centering
\includegraphics[width=0.97\textwidth]{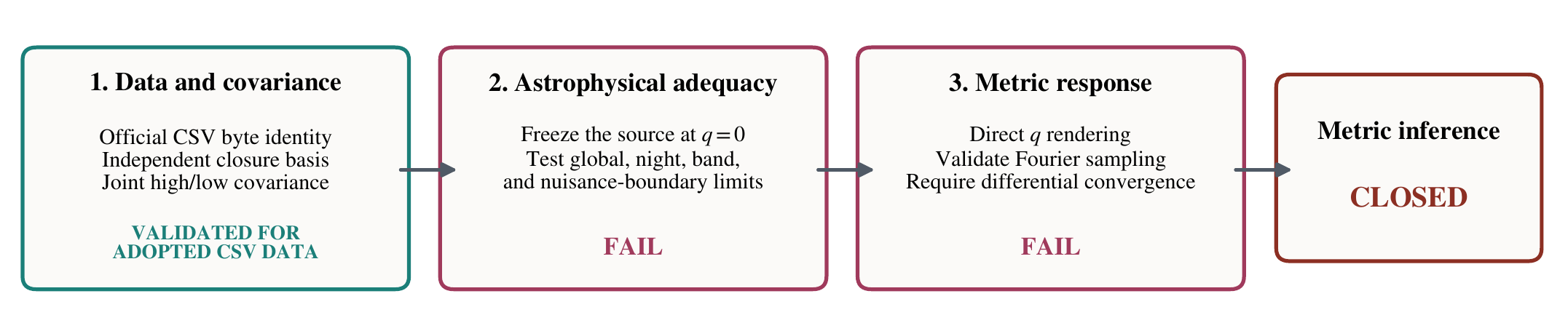}
\caption{The three safeguards used in this work. The official origin and byte identity of the eight adopted CSV inputs are verified, and the independent closure basis and covariance kernel pass the tests used here. This does not mean that the CSV products were independently regenerated from UVFITS. The frozen Kerr source fails the absolute adequacy safeguard, and the differential tidal charge response fails the numerical convergence safeguard. All three safeguards must pass before a deformation posterior is scientifically justified. Metric inference therefore remains closed.}
\label{fig:safeguards}
\end{figure}

This work brings these points together in one conservative analysis of the public 2017 M87* closure data. We first normalize explicit dark matter profiles and show that realistic dark matter gravity is a null control for the observables used here. We then test a frozen Kerr source with an independent closure likelihood that includes covariance. Finally, we test the numerical convergence of the tidal charge response itself. The analysis uses three safeguards, shown in Fig.~\ref{fig:safeguards}:
\begin{enumerate}
\item the data representation and covariance must be validated;
\item the undeformed source and calibration model must pass an absolute adequacy test;
\item the differential response to the deformation must pass an independent convergence test.
\end{enumerate}
A metric posterior is scientifically justified only when all three safeguards pass.

The rest of the paper is organized as follows. Section~\ref{sec:geometry} defines the tidal charge branch and the M87* dark matter controls. Section~\ref{sec:source} describes the frozen source model and direct ray tracing. Section~\ref{sec:data} gives the closure data and likelihood. Section~\ref{sec:validation} presents the synthetic validation. Section~\ref{sec:adequacy} applies the absolute adequacy safeguard to the real data and locates the mismatch. Section~\ref{sec:false} shows how a false deformation signal can appear when this safeguard is ignored. Section~\ref{sec:numerical} tests the numerical response and separates image convergence from metric response convergence. Section~\ref{sec:discussion} discusses the wider meaning, and Sec.~\ref{sec:conclusions} gives the conclusions.
\section{Geometry and the dark matter control}\label{sec:geometry}

\subsection{Common radial form}

We use a separable Kerr-like line element written through
\begin{equation}
\Delta(r)=r^2-2m(r)r+a^2+\mathcal{B}(r),
\qquad
\Sig=r^2+a^2\cos^2\theta,
\label{eq:delta}
\end{equation}
with
\begin{align}
\dd s^2={}&-\left[1-\frac{2m(r)r-\mathcal{B}(r)}{\Sig}\right]\dd t^2
-\frac{2a[2m(r)r-\mathcal{B}(r)]\sin^2\theta}{\Sig}\dd t\,\dd\phi
+\frac{\Sig}{\Delta}\dd r^2+\Sig\dd\theta^2 \nonumber\\
&+\frac{[(r^2+a^2)^2-a^2\Delta\sin^2\theta]\sin^2\theta}{\Sig}\dd\phi^2.
\label{eq:metric}
\end{align}
The Kerr limit is $m(r)=M$ and $\mathcal{B}=0$. For the tidal charge branch,
\begin{equation}
 m(r)=M,\qquad \mathcal{B}(r)=q,
\qquad \Delta_q=r^2-2Mr+a^2+q.
\label{eq:tidal}
\end{equation}
This has the Kerr--Newman algebraic form, but $q$ is not an electromagnetic charge. It is an effective projected Weyl contribution and may be negative \cite{Dadhich2000,Aliev2005,Maartens2010}. The horizon condition is
\begin{equation}
 r_\pm=M\pm\sqrt{M^2-a^2-q},
\qquad a^2+q\leq M^2.
\end{equation}
We study the one-sided branch $q\leq0$ and report $q/M^2$.

For the dark matter control,
\begin{equation}
 m(r)=M+M_{\DM}(r),\qquad \mathcal{B}=0,
\qquad
M_{\DM}(r)=4\pi\int_{r_{\rm min}}^r \rho_{\DM}(r')r'^2\,\dd r'.
\label{eq:radialmass}
\end{equation}
The lower limit is zero for the smooth halo extrapolations and is set by the stated capture prescription for the spike controls. The null geodesic equations remain separable in the adopted phenomenological geometry. Spherical photon orbits satisfy $\Rpot=\Rpot'=0$, where
\begin{equation}
\Rpot(r)=\left[(r^2+a^2)E-aL_z\right]^2
-\Delta\left[\Qcar+(L_z-aE)^2\right].
\end{equation}
These relations provide a simple setting for controlled optical comparisons \cite{Bardeen1973,Carter1968,Chandrasekhar1983,Grenzebach2014,Grenzebach2015,PerlickTsupko2022}.

\subsection{Why this is a dark matter control, not an exact rotating halo}

A radial mass function in a rotating metric is not automatically a cold dark matter solution. The Einstein tensor gives an anisotropic effective source. In the principal frame, the validated density and radial pressure are
\begin{equation}
 \rho=\frac{r^2m'(r)}{4\pi\Sig^2},
 \qquad p_{\parallel}=-\rho.
\label{eq:rho}
\end{equation}
The transverse stresses are generally nonzero and depend on $m'$, $m''$, $r$, and $\theta$. We therefore call this branch a \emph{dark matter environmental control}: the enclosed mass is taken from explicit dark matter profiles, but the rotating radial-mass metric is used only as a phenomenological optical response. It is not presented as an exact rotating pressureless dark matter solution. An independent Einstein-tensor check gave a maximum relative discrepancy of about $2.1\times10^{-5}$.

\subsection{Explicit M87* profiles and normalization}

We use the M87* benchmark of Lacroix, Boehm, and Silk \cite{Lacroix2015}, with $M_{\rm BH}=6.4\times10^9M_\odot$, Schwarzschild radius $R_s=6.0\times10^{-4}\,{\rm pc}$, halo scale $r_0=20\,{\rm kpc}$, inner NFW slope $\gamma=1$, $\rho_0=2.5\,{\rm GeV\,cm^{-3}}$, and $\alpha_\gamma=0.1$. The smooth controls are
\begin{align}
\rho_{\rm NFW}(r)&=\frac{\rho_0}{(r/r_0)(1+r/r_0)^2},\label{eq:nfw}\\
\rho_{\rm Ein}(r)&=\rho_{-2}\exp\left[-\frac{2}{\alpha_E}\left\{\left(\frac{r}{r_{-2}}\right)^{\alpha_E}-1\right\}\right],\label{eq:einasto}
\end{align}
where $\alpha_E=0.18$, $r_{-2}=r_0$, and $\rho_{-2}=\rho_{\rm NFW}(r_0)$. The Einasto curve is used only as a matched comparison. It is not an independent fit to the inner M87* halo.

The spike radius, normalization, and adiabatic slope are
\begin{align}
R_{\rm sp}&=\alpha_\gamma r_0\left(\frac{M_{\rm BH}}{\rho_0r_0^3}\right)^{1/(3-\gamma)},\nonumber\\
\rho_R&=\rho_0\left(\frac{R_{\rm sp}}{r_0}\right)^{-\gamma},
\qquad
\gamma_{\rm sp}=\frac{9-2\gamma}{4-\gamma}=\frac{7}{3},\label{eq:spikepars}\\
\rho_{\rm sp}(r)&=\rho_R\,g(r)\left(\frac{R_{\rm sp}}{r}\right)^{\gamma_{\rm sp}}.\label{eq:spike}
\end{align}
This gives $R_{\rm sp}=220.45\,{\rm pc}$ and $\rho_R=226.81\,{\rm GeV\,cm^{-3}}$. We consider an uncut $\gamma_{\rm sp}=7/3$ envelope, the capture suppressed choice $g(r)=(1-4R_s/r)^3$ for $r>4R_s$ and zero below, and a heated crest with $\gamma_{\rm sp}=3/2$ and the same capture factor \cite{GondoloSilk1999,Merritt2004,Merritt2007,Merritt2002,Lacroix2015}. The uncut case is intentionally optimistic. It is not our preferred physical profile for M87*.

\begin{table}[t]
\caption{M87* dark matter controls. Smooth-halo values are formal inward extrapolations. The rendered upper-envelope control is the normalization used in the validated ray tracing comparison.}
\label{tab:dmprofiles}
\centering
\begin{tabular}{@{}lcc@{}}
\toprule
Profile & Near-horizon prescription & $M_{\rm DM}(<10M)/M_{\rm BH}$\\
\midrule
NFW & smooth extrapolation & $1.16\times10^{-11}$\\
Matched Einasto, $\alpha_E=0.18$ & smooth extrapolation & $1.05\times10^{-14}$\\
Adiabatic spike, $\gamma_{\rm sp}=7/3$ & uncut upper envelope & $7.07\times10^{-5}$\\
Adiabatic spike, $\gamma_{\rm sp}=7/3$ & capture suppressed & $3.31\times10^{-8}$\\
Heated crest, $\gamma_{\rm sp}=3/2$ & capture suppressed & $2.81\times10^{-12}$\\
Rendered upper-envelope control & direct optical calculation & $7.33\times10^{-5}$\\
Old extreme proxy & stress test only & $2.50\times10^{-2}$\\
\bottomrule
\end{tabular}
\end{table}

\begin{figure}[t]
\centering
\includegraphics[width=0.98\textwidth]{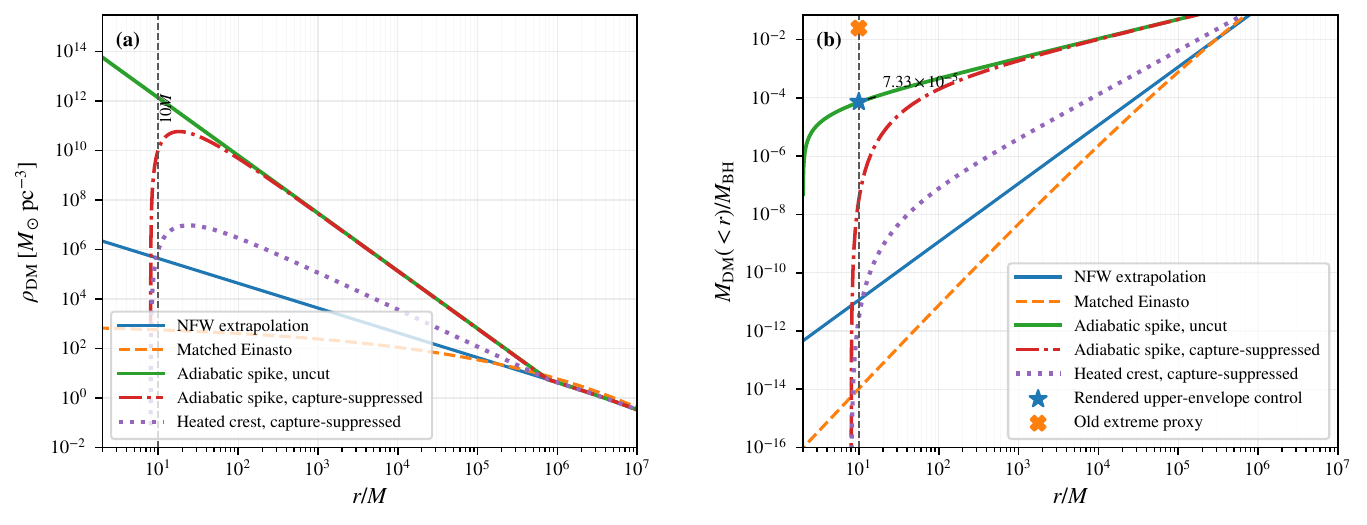}
\caption{Explicit M87* dark matter controls. (a) Density profiles for the NFW and matched Einasto extrapolations, the deliberately optimistic uncut adiabatic spike, the capture suppressed adiabatic spike, and the heated crest. (b) Corresponding enclosed mass fractions. The star marks the upper-envelope normalization used in the direct dark matter optical control, and the cross marks the superseded extreme proxy. The dashed vertical line gives $r=10M$.}
\label{fig:dmcontrol}
\end{figure}

Figure~\ref{fig:dmcontrol} shows the density and enclosed-mass hierarchy. The smooth NFW and Einasto extrapolations are negligible. A capture suppressed canonical spike also lies more than three orders of magnitude below the deliberately optimistic rendered control at $10M$. The old proxy value of $0.025$ is about 340 times larger. We keep it only to show how extreme the first benchmark was.

In the common semi analytic transfer calculation, the rendered upper-envelope dark matter control gave an image total-variation distance of $2.65\times10^{-6}$ from Kerr and an RMS normalized-visibility residual of $4.9\times10^{-7}$. For comparison, $q/M^2=-0.02$ gave approximately $8.35\times10^{-3}$ and $1.0\times10^{-3}$. Only the upper envelope normalization was ray traced directly. The other profiles set the physically allowed enclosed mass scale. Since the capture suppressed and smooth profiles contain orders of magnitude less mass over the same near horizon region, the directly rendered values provide a deliberately conservative optical control rather than separate image predictions for every profile. Dark matter is therefore discussed explicitly and remains physically central to the motivation, but it does not act as a fitted competitor to tidal charge in the present M87* likelihood.

\section{Frozen source model and direct ray tracing}\label{sec:source}

\subsection{Why the source is frozen at Kerr}

The source family was selected and refined only at $q=0$. This avoids changing the source after the deformation result is known. The final shared geometry is
\begin{equation}
 a/M=-0.94,\qquad i=22^\circ.
\label{eq:geomfixed}
\end{equation}
The sign of $a$ follows the screen-orientation convention used in the ray tracer. Eight source assignments describe the early and late segments of the four observing nights. They are built from four unique compact-image states plus bounded analytic nuisance freedom. The source nodes, segment assignments, parity choices, parameter bounds, covariance scenarios, and adequacy limits are frozen throughout the metric test.

The compact source is a dual-cone plus equatorial surface model. Its radial profile is
\begin{equation}
 J(r;R,p_1,p_2)=\frac{(r/R)^{p_1}}{1+(r/R)^{p_1+p_2}}.
\label{eq:radialprofile}
\end{equation}
At each retained surface crossing, the weight has the schematic form
\begin{equation}
 W\propto J(r)\,\delta^{3+\sigma}\,\ell_{\rm path}\,
\sin^{1+\sigma}\!\zeta,
\label{eq:surfaceweight}
\end{equation}
where $\delta$ is the fluid-frame redshift factor, $\ell_{\rm path}$ is a capped thin-surface path-length proxy, and $\zeta$ is the angle between the photon direction and the magnetic direction. The cone and equatorial contributions are mixed as
\begin{equation}
 I=(1-r_J)I_{\rm cone}+r_J I_{\rm eq}.
\end{equation}
This model is useful for a controlled inference test, but it is not a volumetric general relativistic radiative transfer calculation of the kind used in modern code comparisons and EHT modeling studies \cite{Porth2019,Gold2020,Bronzwaer2020,Wong2022}. That limitation becomes important in Sec.~\ref{sec:numerical}.

\subsection{Compact and extended visibility components}

The normalized Fourier transform of the compact image is combined with an offset elliptical Gaussian,
\begin{equation}
 V(u,v)=f_cV_{\rm compact}(u,v)+(1-f_c)V_{\rm ext}(u,v).
\label{eq:visibility}
\end{equation}
The eight bounded nuisance parameters are the angular scale, compact position angle, compact flux fraction, extended major-axis FWHM, extended axis ratio, extended position angle, offset radius, and offset position angle. These parameters are allowed to refit within their frozen bounds for each covariance scenario. The source images and the source family itself are not reopened.

\begin{figure}[t]
\centering
\includegraphics[width=0.96\textwidth]{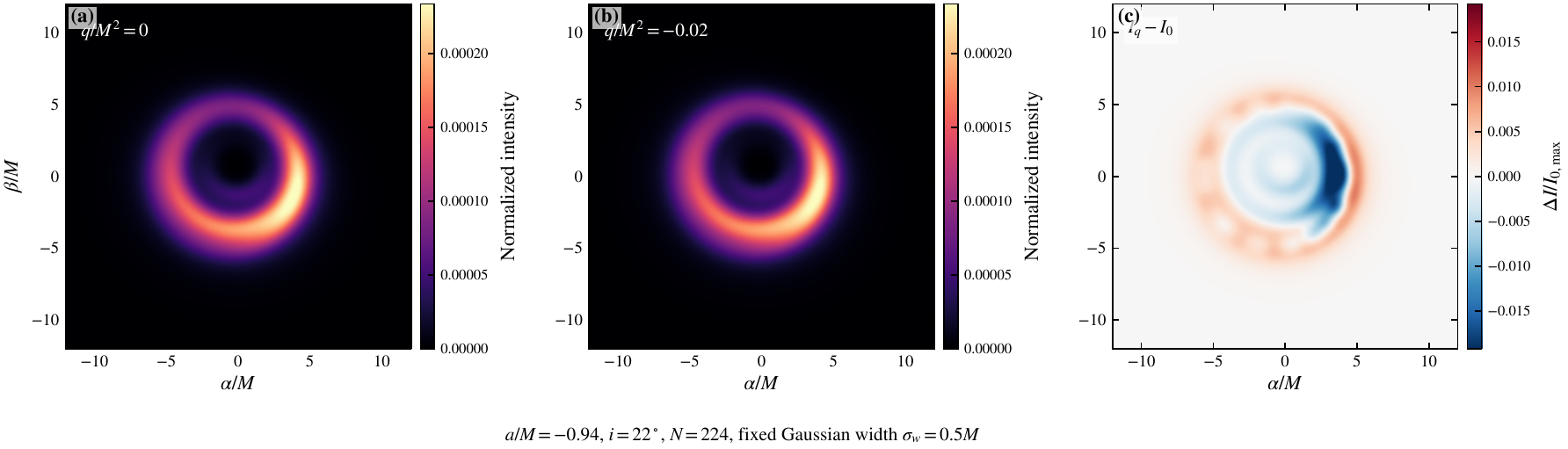}
\caption{Illustrative frozen-source images at $a/M=-0.94$ and $i=22^\circ$. The first two panels show Kerr and $q/M^2=-0.02$ after the same fixed screen-plane smoothing. The third panel shows their difference. The images are normalized separately. This figure is illustrative only and is not used to claim a measurement of $q$.}
\label{fig:images}
\end{figure}

Figure~\ref{fig:images} gives one illustrative compact state. The two images are visually close, while their small difference carries the deformation information. This is exactly why response convergence must be tested more strictly than image appearance.

\subsection{Direct charge libraries}

The one-sided charge grid is
\begin{equation}
q/M^2=0,-0.0005,-0.001,-0.0025,-0.005,-0.0075,
-0.01,-0.015,-0.02,-0.03,-0.04,-0.06.
\label{eq:qgrid}
\end{equation}
The first direct bank used $N=96$. A full $N=128$ bank and representative $N=160$ images revealed a resolution problem. The production rebuild then used $N=192$, 224, and 256 on a fixed $24M$ field of view. The final production library contains 92 direct images. All rays ended in accepted capture, escape, or completed-crossing states, and the maximum constraint projection was $2.0\times10^{-5}$.

The Fourier transform used a padding factor of 12, a square-pixel top-hat correction, and interpolation on the actual M87* coverage. Direct discrete Fourier sums gave a maximum complex discrepancy of $9.38\times10^{-4}$, below the fixed $10^{-3}$ gate. Thus, the final instability is not explained by unresolved rays or ordinary FFT interpolation error.

\section{Public closure data and covariance-aware likelihood}\label{sec:data}

\subsection{Data representation}

We use the eight public 2017 M87* Stokes-$I$ CSV products: four observing days (95, 96, 100, and 101) and two frequency bands. The calibration and data-processing context is described by the EHT Collaboration \cite{EHT2019III}. The analysis contains 1372 independent closure observables organized into 136 covariance blocks and 72 scan blocks.

For baseline phases $\bm{\phi}$ and log amplitudes $\bm{a}$, independent closure bases are built as
\begin{equation}
 \bm{\psi}=\mathbf{A}_{\rm cp}\bm{\phi},
 \qquad
 \bm{c}=\mathbf{A}_{\rm ca}\bm{a}.
\label{eq:closurebasis}
\end{equation}
The incidence matrices are reduced to independent rows. Their covariances follow by linear propagation. This avoids counting redundant closures as independent data and follows the closure statistics framework developed for interferometric closure quantities \cite{Jennison1958,Thompson2017,Blackburn2020}. Raw visibility amplitudes are not added as a separate likelihood block.

The official EHT data release provides both representations under data product 2019-D01-01 and states that its CSV products were derived from the released UVFITS files \cite{EHT2019DataRelease}. We downloaded all eight CSV files again and compared them byte by byte with the local inputs used in this analysis. Every pair has the same file size and SHA-256 digest. This verifies the origin and integrity of the adopted CSV inputs. We do not claim that we independently regenerated the CSV files from the binary UVFITS data.

\subsection{High/low-band covariance}

For each simultaneous high/low scan and closure type, the joint covariance is
\begin{equation}
 \Cmat_{\rm joint}=
 \begin{pmatrix}
 \Cmat_{\rm hi} & \mathbf{X}\\
 \mathbf{X}^{\mathsf T} & \Cmat_{\rm lo}
 \end{pmatrix}.
\label{eq:jointcov}
\end{equation}
The diagonal blocks preserve the validated within-band covariance exactly. The cross block propagates a fixed fraction of the official one-degree phase floor or two-percent amplitude floor through the common closure-incidence structure. The empirical extra scatter inferred from simultaneous high-minus-low differences remains band specific:
\begin{equation}
 s_{\rm cp}=6.0966^\circ,
 \qquad s_{\rm lca}=0.034284.
\end{equation}
A common-mode term cancels from a band difference, so its fraction cannot be estimated from those differences. We therefore predeclare three sensitivity cases:
\begin{center}
\begin{tabular}{lcc}
\toprule
Scenario & Common fraction of official floor variance & Role\\
\midrule
C0 & 0 & independent-band limit\\
C50 & 0.5 & central sensitivity case\\
C100 & 1 & maximally common official-floor limit\\
\bottomrule
\end{tabular}
\end{center}
The common fraction is never optimized against the real observations.

The score is
\begin{equation}
 \chi^2=\rvec^{\mathsf T}\Cmat^{-1}\rvec,
 \qquad \rvec=\mathbf{d}-\muv(\bm{\vartheta},q),
\label{eq:chi2}
\end{equation}
where $\bm{\vartheta}$ denotes the bounded nuisance parameters. Marginal band diagnostics use the unchanged within-band covariance and are not added as independent terms to the correlated joint statistic.

\subsection{Predeclared adequacy limits}

The limits were fixed before the deformation was opened:
\begin{align}
\chi^2_{\rm global}/N&\leq1.5,\nonumber\\
\chi^2_{\rm night}/N&\leq2.0\quad\text{for every night},\nonumber\\
\chi^2_{\rm band}/N&\leq2.0\quad\text{for every band},
\label{eq:gates}
\end{align}
with no unexplained nuisance-boundary solutions. All C0, C50, and C100 cases must pass. These are practical safeguards, not universal constants. They are fixed model adequacy checks and are not tuned after seeing the result \cite{GelmanMengStern1996,BayarriBerger2000,Andrae2010}. The global value 1.5 allows a 50\% excess above the calibrated Kerr expectation. In the 5000 synthetic draws, the standard deviation of $\chi^2/N$ is only about 0.038. The night and band value 2.0 is intentionally looser and is used only to identify a severe local mismatch. The limits were fixed before the deformation was opened and are not changed after seeing its direction.

\begin{table}[t]
\caption{Frozen inference state.}
\label{tab:frozen}
\centering
\begin{tabular}{@{}l@{\hspace{1em}}l@{}}
\toprule
Item & \parbox[t]{0.67\textwidth}{Frozen choice}\\
\midrule
Geometry & \parbox[t]{0.67\textwidth}{$a/M=-0.94$, $i=22^\circ$; deformation closed during source selection}\\[2pt]
Source & \parbox[t]{0.67\textwidth}{Dual-cone/equatorial surface family; eight day-segment assignments; four unique compact states}\\[2pt]
Nuisance family & \parbox[t]{0.67\textwidth}{Eight bounded analytic parameters in the compact plus extended visibility model}\\[2pt]
Covariance & \parbox[t]{0.67\textwidth}{Independent closure basis; C0/C50/C100 common-floor sensitivity triad}\\[2pt]
Adequacy & \parbox[t]{0.67\textwidth}{Global $\chi^2/N\leq1.5$; every night and band $\leq2.0$; no unexplained boundaries}\\[2pt]
Metric branch & \parbox[t]{0.67\textwidth}{$q/M^2\leq0$; direct rendering on the fixed grid in Eq.~\eqref{eq:qgrid}}\\[2pt]
Numerical gate & \parbox[t]{0.67\textwidth}{Stable differential closure response and nuisance-projected $\Delta\chi^2$ across successive production resolutions}\\
\bottomrule
\end{tabular}
\end{table}

\section{Synthetic validation of the statistical kernel}\label{sec:validation}

Five thousand Kerr-only draws were generated for each covariance scenario over the actual block structure. The tests include the mean and variance of the normalized statistic $\chi^2/N$, central 95\% coverage, recovery of a one-percent global angular-scale perturbation, 68\% and 95\% interval coverage, and the nominal five-percent null false-positive rate. The scale perturbation is only a nuisance-recovery test; it is not used as a proxy for tidal charge.

The results are given in Table~\ref{tab:synthetic}. The mean normalized statistic is within $4\times10^{-4}$ of one. The central 95\% coverage lies between 0.9472 and 0.9546, and the null false-positive rate lies between 0.0482 and 0.0552. Positive definiteness and whitening tests also pass. The literal C0 fixed-parameter calculation reproduces the earlier independent-band checkpoint to machine precision.

\begin{table}[t]
\caption{Synthetic Kerr validation and real data frozen-source adequacy.}
\label{tab:synthetic}
\centering
\begin{tabular}{lcccccc}
\toprule
Scenario & Mean synthetic $\chi^2/N$ & 95\% coverage & False positive & Real global $\chi^2/N$ & Worst night & Worst band\\
\midrule
C0   & 0.9996 & 0.9490 & 0.0482 & 1.7583 & 1.9831 & 2.0216\\
C50  & 0.9997 & 0.9472 & 0.0552 & 1.7206 & 1.9559 & 2.0216\\
C100 & 1.0003 & 0.9546 & 0.0514 & 1.6873 & 1.9335 & 2.0216\\
\bottomrule
\end{tabular}
\end{table}

\begin{figure}[t]
\centering
\includegraphics[width=0.92\textwidth]{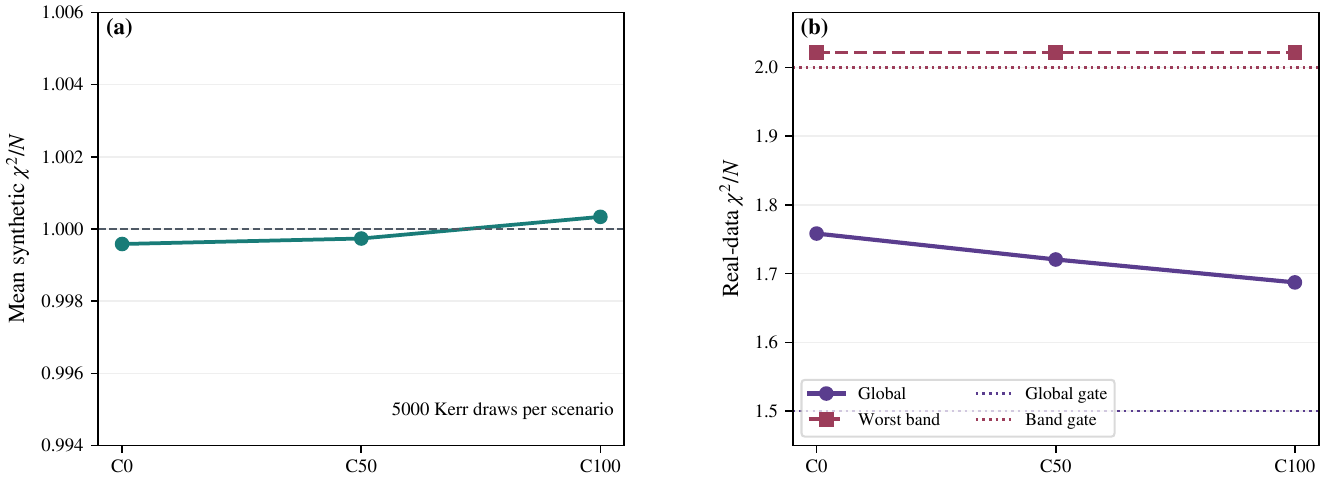}
\caption{Statistical validation and absolute adequacy. (a) The synthetic Kerr draws give the expected mean normalized statistic. (b) The frozen real data Kerr fits fail the fixed global gate in all covariance scenarios. The worst simultaneous band is also above its gate.}
\label{fig:validation}
\end{figure}

The left panel of Fig.~\ref{fig:validation} shows the synthetic result. The right panel shows the very different conclusion for the real data. This separation is central: a correct statistical kernel does not imply that the source model is adequate.

A modest threshold-sensitivity check is given in Table~\ref{tab:threshold}. It does not rescue the source model under the predeclared all-scenario rule. All three covariance cases fail at 1.5 and 1.6. At 1.7 the C100 case alone passes, and at 1.75 only C0 still fails. The conclusion becomes a universal pass only when the global limit is relaxed to 1.8, which would then accept the observed 69--76\% excesses over the synthetic expectation. This check is descriptive and is not used to redefine the gate.

\begin{table}[t]
\caption{Sensitivity of the global adequacy verdict to a hypothetical common threshold. ``Pass all'' requires C0, C50, and C100 to pass together.}
\label{tab:threshold}
\centering
\begin{tabular}{@{}ccccc@{}}
\toprule
Threshold & C0 & C50 & C100 & Pass all\\
\midrule
1.50 & fail & fail & fail & no\\
1.60 & fail & fail & fail & no\\
1.70 & fail & fail & pass & no\\
1.75 & fail & pass & pass & no\\
1.80 & pass & pass & pass & yes\\
\bottomrule
\end{tabular}
\end{table}

The one-sided $q\leq0$ boundary statistic was separately checked with 200,000 draws. The Chernoff mixture gives the five-percent threshold $\Delta\chi^2=2.705543$ \cite{Chernoff1954}. The measured false-positive rates range from 4.923\% to 5.018\%. Covariance-misspecification tests across all true/assumed C0/C50/C100 pairs give 4.62--5.44\%. These tests validate the likelihood conditional on a correct and numerically stable mean model.

\section{The deformation-blind adequacy safeguard}\label{sec:adequacy}

\subsection{Global, night, and band tests}

The real data results in Table~\ref{tab:synthetic} fail the global limit for all three covariance cases. Including more common high/low covariance lowers the statistic from 1.7583 to 1.6873, but it does not bring it below 1.5. The worst night remains below 2.0, while the worst band remains at 2.0216. Two or three nuisance parameters also reach frozen boundaries, depending on the covariance case.

These values are not close enough to justify opening the deformation. Under C100, the total $\chi^2$ would need to fall by 11.10\% to reach the global threshold. A deformation could produce such a fall, but that would not show that the spacetime is different. It would only show that the extra direction is able to absorb part of the residual.

\subsection{Residual localization}

The exact scan statistic is evaluated block by block,
\begin{equation}
 \chi^2_s=\rvec_s^{\mathsf T}\Cmat_s^{-1}\rvec_s.
\end{equation}
The scan blocks are additive because they partition independent covariance blocks. Station and baseline-scale views are kept only as diagnostics and are not counted again.

The largest structures include the late part of day 101, several late scans on day 96, and late scans on day 100. They span closure phase and log closure amplitude and appear across several stations and baseline scales. The rank ordering is stable across C0, C50, and C100.

\begin{figure}[t]
\centering
\includegraphics[width=0.94\textwidth]{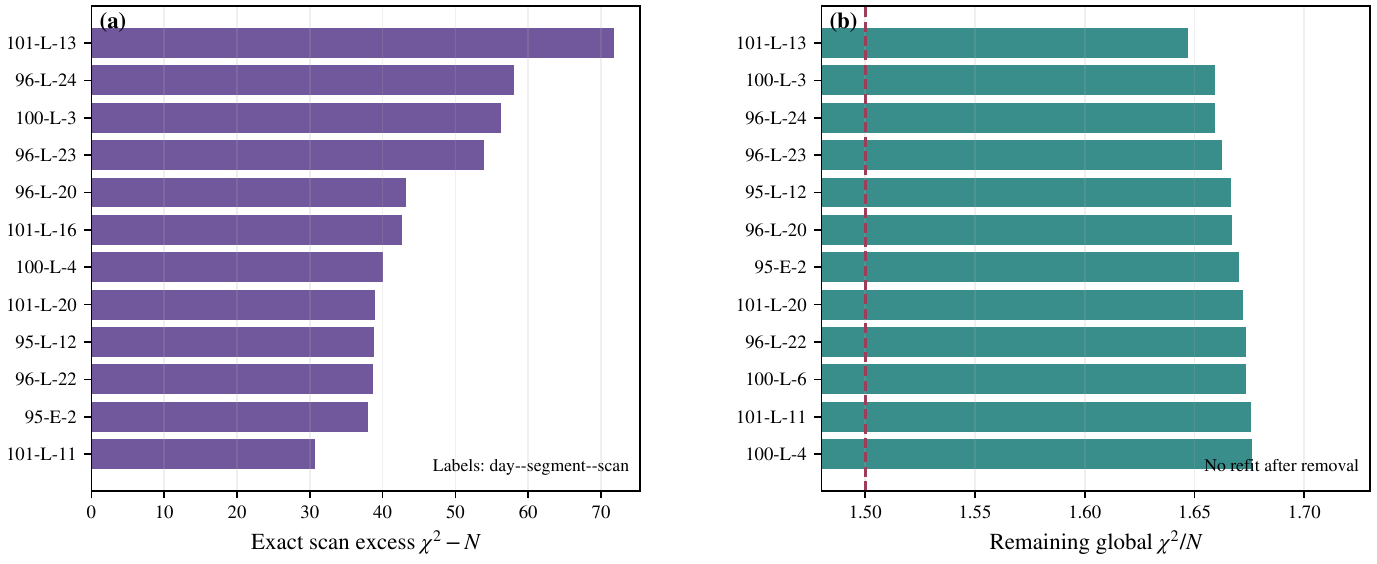}
\caption{Deformation-blind residual localization for C100. (a) Largest exact scan excesses. Labels give day, early/late segment, and scan number. (b) Remaining global statistic after removing one scan without refitting. Even the best removal remains above the fixed global gate.}
\label{fig:localization}
\end{figure}

Figure~\ref{fig:localization}(a) shows the largest exact C100 scan excesses. The mismatch is concentrated, but it is not a single isolated point. To test this directly, every scan is removed once without refitting. The best C100 result is obtained after removing the most influential scan, but the remaining score is still
\begin{equation}
 \chi^2/N=1.6472>1.5.
\end{equation}
Figure~\ref{fig:localization}(b) shows the best removals. No single-scan deletion repairs the model.

The scientifically honest conclusion is therefore simple: the frozen semi analytic source is not an adequate absolute description of the public closure data. Reopening the source bank after seeing the residual, tuning the covariance fraction, deleting several scans, or weakening the threshold would destroy the predeclared test. The deformation must remain closed.

\section{How a false deformation signal appears}\label{sec:false}

The purpose of this section is not to estimate $q$. It is to show what happens if the adequacy rule is ignored.

\subsection{Nuisance-projected deformation direction}

Let the whitened local deformation tangent be
\begin{equation}
 \mathbf{d}_q=\Cmat^{-1/2}\frac{\partial\muv}{\partial q}\bigg|_{q=0},
\end{equation}
with the nuisance Jacobian
\begin{equation}
 \Jmat=\Cmat^{-1/2}\frac{\partial\muv}{\partial\bm{\vartheta}}\bigg|_{q=0}.
\end{equation}
A singular-value decomposition defines the projector $\Pperp$ orthogonal to the local nuisance tangent. The local information is
\begin{equation}
 \mathcal{I}_q=\|\Pperp\mathbf{d}_q\|^2,
 \qquad \sigma_q=\mathcal{I}_q^{-1/2}.
\label{eq:fisher}
\end{equation}
For a directly rendered finite charge,
\begin{equation}
 \Delta\chi^2(q)=\left\|\Pperp\Cmat^{-1/2}
 [\muv(q)-\muv(0)]\right\|^2.
\label{eq:dchi}
\end{equation}
This treatment is exact in the rendered finite-$q$ displacement but local in nuisance profiling.

At the early $N=96$ resolution, the inadequate real residual projected strongly onto the negative-$q$ tangent. The diagnostic scores were $z=-4.73$, $-4.89$, and $-5.19$ for C0, C50, and C100. If read without the adequacy safeguard, this would look like strong evidence for a deformation. It is not. The base source model already fails, and the score only measures alignment between the residual and one extra model direction.

\subsection{Resolution-dependent sign}

The production audit made the warning stronger. After the renderer was rebuilt and fixed Gaussian widths were applied, the same residual projection changed sign between $N=192$ and 224 for every tested width and covariance case. For example, at $\sigma_w=0.25M$ under C100,
\begin{equation}
 z=-5.45\quad(N=192),
 \qquad z=+5.34\quad(N=224).
\end{equation}
The robust statement is therefore not that the mismatch prefers negative tidal charge. The robust statement is that a misspecified source can generate a large formal deformation score, while even the inferred direction can depend on discretization.

\begin{figure}[t]
\centering
\includegraphics[width=0.92\textwidth]{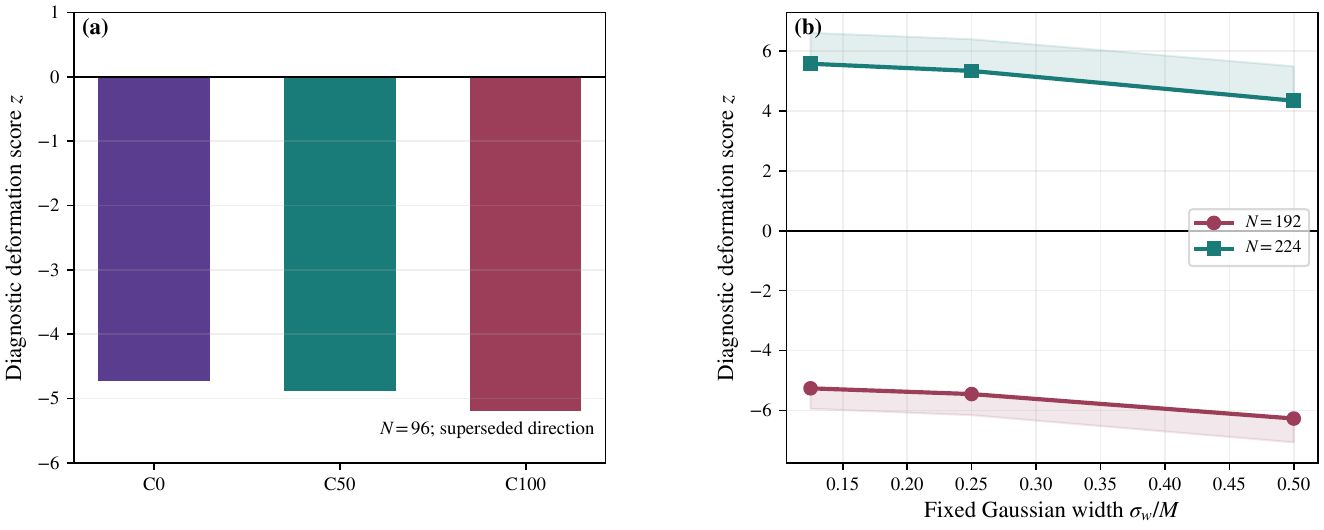}
\caption{False deformation diagnostics. (a) At $N=96$, the inadequate residual projects strongly onto the negative-$q$ direction. This direction is superseded by the later resolution audit. (b) At production resolutions, the score changes sign between $N=192$ and 224. Lines show C100, while shaded bands span C0--C100. None of these scores is a measurement of $q$.}
\label{fig:false}
\end{figure}

This is a concrete example of a general misspecification problem \cite{White1982,KennedyOHagan2001,KleijnVanDerVaart2012}. A likelihood can be internally correct, yet its parameter estimate can be physically misleading because the mean model is wrong. In strong-gravity work, the danger is larger because a deformation parameter often changes the image in a smooth way that can resemble ordinary source structure.

\section{The numerical response safeguard}\label{sec:numerical}

\subsection{Why the metric response is harder than the image}

A small metric deformation is inferred from a difference between two images or two closure vectors. Let $I_N(q)$ be the image at resolution $N$. Absolute image convergence asks whether
\begin{equation}
 I_{N_2}(0)\simeq I_{N_1}(0).
\end{equation}
Metric response convergence instead asks whether
\begin{equation}
 [I_{N_2}(q)-I_{N_2}(0)]
 \simeq
 [I_{N_1}(q)-I_{N_1}(0)].
\label{eq:responseconv}
\end{equation}
The second difference can be much smaller than either image. A small discretization change that is harmless for the image can therefore be large compared with the deformation signal.

Following standard verification practice, the production convergence limits were fixed before the final audit \cite{Celik2008,Roy2005,OberkampfRoy2010}: at most 15\% change in representative finite-$q$ $\Delta\chi^2$, 10\% in local $\sigma(q)$, 15\% in power thresholds, 10\% median projected-vector difference, 25\% at the 90th percentile, and a median vector cosine of at least 0.995. The FFT/direct-sum error had to remain below $10^{-3}$.

\subsection{Production result}

The direct $N=192$, 224, and 256 libraries fail the response limits. The largest $N=192\rightarrow224$ finite-$q$ $\Delta\chi^2$ change is 20.57\%. For the shared C100 nodes between $N=224$ and 256, the maximum change is 61.88\%. The local $\sigma(q)$ changes by 35.68\%, and the median nuisance-projected response vector changes by 75.13\% with a median cosine of 0.7142.

\begin{figure}[t]
\centering
\includegraphics[width=0.93\textwidth]{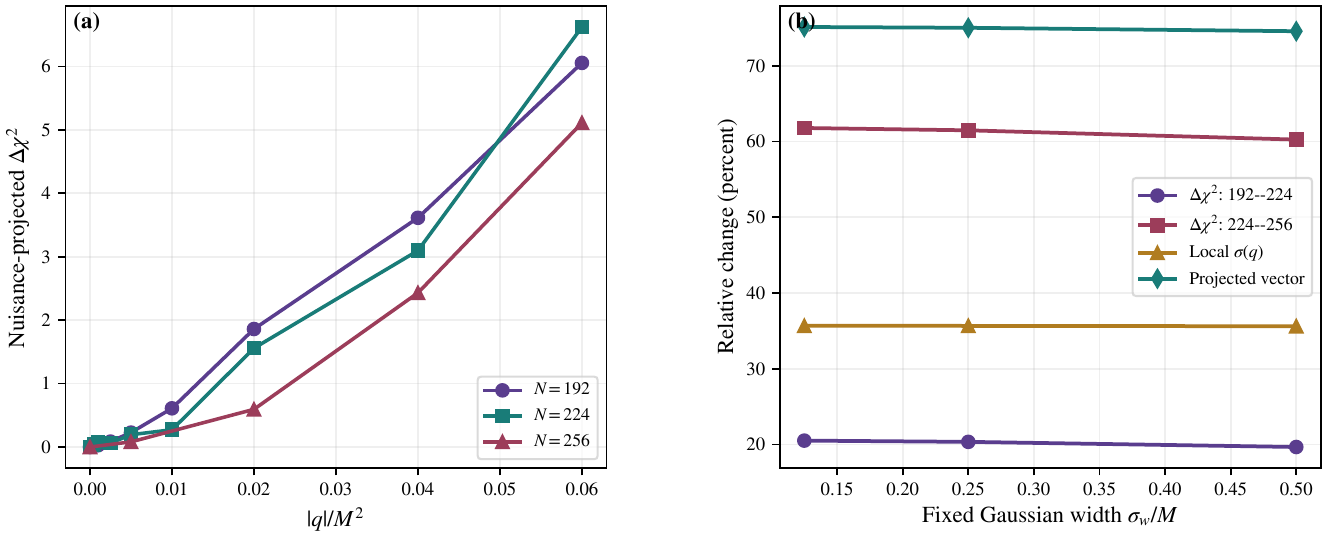}
\caption{Failure of the numerical response safeguard. (a) Direct C100 nuisance-projected separation at the production resolutions. The finite-$q$ response moves significantly under refinement. (b) Maximum relative changes after fixed Gaussian regularization. None of the tested widths makes the response stable.}
\label{fig:response}
\end{figure}

Figure~\ref{fig:response}(a) shows the C100 direct profiles. The curves are smooth at each resolution, but they are not stable between resolutions. This matters more than visual smoothness. The provisional sensitivity values from the lower-resolution banks are therefore diagnostics only and are not quoted as EHT or future-array limits.

\subsection{Pixel integration and fixed-width regularization}

A deterministic $2\times2$ detector-pixel integral was applied to the direct image bank. It did not solve the problem: the maximum effective $96\rightarrow112$ $\Delta\chi^2$ change was 24.55\%, the effective $112\rightarrow128$ C100 change was 60.98\%, and the median projected-vector change was 73.33\%.

We then applied fixed screen-plane Gaussian widths
\begin{equation}
 \sigma_w/M=0.125,\ 0.25,\ 0.50.
\end{equation}
The width was held fixed across charge, source state, resolution, day, segment, and covariance scenario. It was not tuned to the observations. Table~\ref{tab:numerical} summarizes the result. Every response gate fails for every width, while the Fourier gate continues to pass.

\begin{table}[t]
\caption{Production response-convergence audit. Values are maximum or median relative changes, as indicated.}
\label{tab:numerical}
\centering
\begin{tabular}{lcccccc}
\toprule
$\sigma_w/M$ & $\Delta\chi^2$ 192--224 & C100 $\Delta\chi^2$ 224--256 & $\sigma(q)$ & Threshold & Vector difference & Vector cosine\\
\midrule
0.125 & 20.52\% & 61.78\% & 35.67\% & 23.69\% & 75.11\% & 0.7140\\
0.250 & 20.35\% & 61.47\% & 35.66\% & 23.49\% & 75.00\% & 0.7136\\
0.500 & 19.67\% & 60.25\% & 35.61\% & 22.53\% & 74.55\% & 0.7147\\
\bottomrule
\end{tabular}
\end{table}

The weak dependence on $\sigma_w$ shows that the instability is not removed by modest image smoothing. The likely origin is the point-sampled, grazing-enhanced surface architecture itself. Fixed screen smoothing is a numerical sensitivity test, not a physical finite-thickness plasma model.

\subsection{Absolute image convergence versus response convergence}

The final image-space audit places all resolutions on the same physical grid, normalizes the images, applies the same fixed physical smoothing, and compares both the Kerr image and the finite-$q$ response. We use the relative norm
\begin{equation}
 D_2(A,B)=\frac{\|A-B\|_2}{\|A\|_2}
\end{equation}
with the cosine similarity
\begin{equation}
 \mathcal{C}(A,B)=\frac{A\cdot B}{\|A\|_2\|B\|_2}.
\end{equation}

\begin{figure}[t]
\centering
\includegraphics[width=0.93\textwidth]{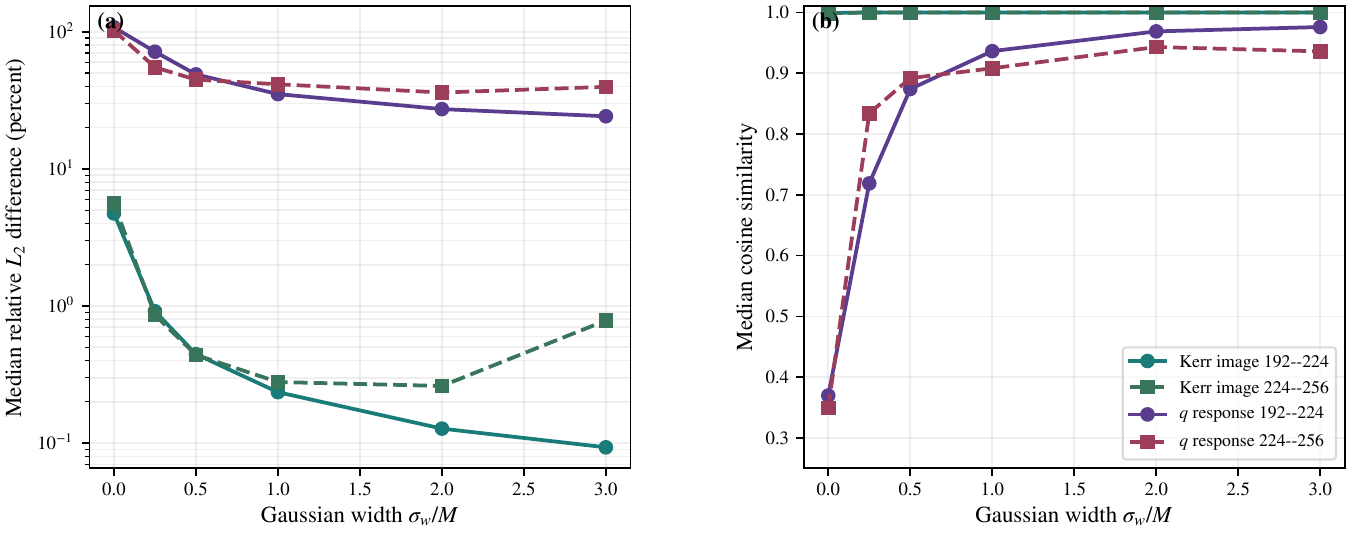}
\caption{Absolute image convergence is not metric response convergence. The Kerr images become very stable after modest fixed smoothing, while the small tidal charge difference remains much less stable. The response must therefore be tested independently of the image.}
\label{fig:image-response}
\end{figure}

At $\sigma_w=0.5M$, the median Kerr-image difference between $N=192$ and 224 is 0.4455\%, with cosine 0.999990. The median $q$-response difference is 48.92\%, with cosine 0.8740. Even at $\sigma_w=1M$, the response difference remains 35.12\%. Figure~\ref{fig:image-response} shows the full trend.

This is the decisive numerical result. A visually stable image is not enough for a small-deformation inference. The response in Eq.~\eqref{eq:responseconv} must pass its own convergence test.

\section{Discussion}\label{sec:discussion}

\subsection{What has passed and what has not}

Several parts of the analysis pass their tests. The independent closure construction, covariance algebra, positive-definiteness checks, synthetic whitening, Kerr recovery, boundary calibration, and Fourier/direct-sum comparison all pass. This helps us locate the actual problem. The main limitations are not ordinary likelihood coding errors or unresolved rays.

For the adopted CSV representation, the data and covariance safeguard passes the tests used here. The eight local CSV inputs are byte identical to fresh downloads from the official release, and the release documents their UVFITS origin. The independent closure construction, covariance algebra, positive definiteness checks, synthetic whitening, Kerr recovery, boundary calibration, and Fourier/direct-sum comparison also pass. We do not claim that the CSV files were independently regenerated from UVFITS. The two failures are scientific: the frozen source model does not pass the absolute real data adequacy test, and the differential tidal charge response does not converge at the tested resolutions and regularizations.

A real data posterior for $q$ would therefore combine an inadequate mean model with an unstable response. A narrow interval or a large Bayes factor could still be computed, but it would not have a defensible physical meaning. We deliberately do not report such quantities.

\subsection{Why the negative result is useful}

A negative inference result can be more useful than an unreliable parameter bound. Here it gives four clear lessons:
\begin{enumerate}
\item a covariance-aware closure likelihood can pass synthetic tests while the source model fails on real data;
\item the mismatch can be distributed over several scans and cannot be repaired by deleting one point;
\item opening a deformation can create a large formal score from source inadequacy;
\item the direction of that score can reverse when the differential response is numerically refined.
\end{enumerate}
These points are relevant beyond tidal charge. They apply to any small deformation whose observable effect can be confused with source structure.

\subsection{Relation to current EHT inference}

The closure covariance and independent-degree problem is well known \cite{Blackburn2020}. The EHT calibration papers also document nontrivial systematics and day-scale evolution \cite{EHT2019III}. Lockhart and Gralla showed that closure-likelihood choices can change geometric ring inferences \cite{LockhartGralla2022}. More recent multi epoch work shows that intrinsic variability and model misspecification can shift orientation parameters in semi analytic fits \cite{Chang2025}, while short-baseline closures can carry information about extended structure \cite{Georgiev2026}.

The dark matter calculation gives a clear physical baseline for this inference test. It shows that NFW, Einasto, and capture suppressed spike gravity cannot explain the structured closure mismatch, although an unrealistically dense proxy can be made visible. The contribution of this work is the combined, predeclared safeguard framework. We do not claim priority for closure covariance, variability, dark matter spikes, or model misspecification separately. The new point is that data validity, source adequacy, and differential response convergence must all be established before a metric posterior is opened. The public M87* example shows why each safeguard is needed.

\subsection{What a future metric analysis requires}

A new attempt to constrain $q$ should not continue by merely increasing the present screen resolution. It should begin with a physically smooth volumetric emissivity and absorptivity model, preferably with a controlled general relativistic radiative transfer calculation informed by GRMHD or a carefully validated semi analytic volume \cite{Gammie2003,Dexter2009,Moscibrodzka2016,Bronzwaer2018,Chael2018,Porth2019,Gold2020,Wong2022}. The new source bank must be selected and frozen at $q=0$. It must pass posterior predictive or absolute fit tests across nights and bands before the deformation is opened.

The metric libraries should then be rendered directly at several production resolutions. Convergence should be checked in the sampled complex visibilities, independent closure vectors, nuisance projected finite-$q$ displacements, and the final likelihood ratio. Comparing images by eye is not enough. Multi epoch data should also be used to separate persistent geometric structure from variable emission. An independent UVFITS to CSV regeneration would be a useful extra check, although the adopted CSV inputs already pass the official origin and byte identity tests.

\subsection{Scope and limitations}

The source model is intentionally compact and transparent. This is useful for a controlled audit, but it is not a final model of M87*. The fixed adequacy limits are practical scientific safeguards rather than universal constants. Another analysis may choose different limits, but they should be declared before the deformation result is known and should be tested with simulations.

The tidal charge branch is used as a simple one parameter deformation. The paper does not test all non-Kerr metrics, and it does not claim that a negative tidal charge is physically favored or disfavored by M87*. The dark matter profiles are prior-predictive controls, not measurements of the inner halo. The smooth NFW and Einasto curves are inward extrapolations, while the uncut spike is intentionally optimistic. Their role is to bracket the gravity of plausible dark matter environments. They are not used to infer dark matter particle properties.

Finally, the analysis is grounded in the official public CSV products. Their provenance is documented by the release, and all eight local inputs are byte-identical to newly downloaded official copies. No claim is made that this work independently regenerated those CSV files from UVFITS.

\section{Conclusions}\label{sec:conclusions}

We tested a rotating tidal charge deformation with the public 2017 M87* closure data, a frozen semi analytic source family, and a covariance-aware independent closure likelihood. Dark matter was treated explicitly through NFW, matched Einasto, canonical-spike, capture suppressed, and heated-crest controls. Even the optimistic rendered normalization $M_{\rm DM}(<10M)/M_{\rm BH}=7.33\times10^{-5}$ produced negligible image and visibility changes. Realistic dark matter gravity therefore cannot explain the structured mismatch found in the closure data.

The closure-statistics kernel passes its synthetic Kerr tests. The real data Kerr source does not. Its global scores are $1.7583$, $1.7206$, and $1.6873$ for C0, C50, and C100, above the fixed limit of $1.5$. The worst band gives $2.0216$. The residual is structured across several scans, and removing the single most influential scan still leaves $1.6472$.

If the deformation is opened despite this failure, the residual creates a large formal deformation score. The apparent direction is not stable: it changes sign between $N=192$ and 224. The direct finite-$q$ response also fails the production convergence tests after completed-ray validation, direct Fourier checks, pixel integration, and fixed-width smoothing.

The clearest result is that image convergence and metric response convergence are different requirements. At $\sigma_w=0.5M$, the absolute Kerr image changes by about 0.45\% between $N=192$ and 224, while the tidal charge response changes by about 49\%. A stable looking image can therefore support an unstable deformation inference.

We do not report a tidal charge posterior, upper bound, Bayes factor, or detection threshold. The main conclusion is the three safeguard rule:
\begin{quote}
A spacetime deformation should not be inferred until the adopted data and covariance are validated, the undeformed source model passes an absolute adequacy test, and the differential deformation response passes an independent numerical convergence test.
\end{quote}
For the adopted CSV representation, the data and covariance safeguard passes. The source adequacy and metric response safeguards fail. Metric inference must therefore remain closed.

\begin{acknowledgments}
The author acknowledges financial support from ANID through FONDECYT Postdoctoral Grant No.~3260029.
\end{acknowledgments}

\section*{Data and code availability}
The analysis uses the public 2017 M87* Stokes-$I$ CSV products from EHT data product 2019-D01-01 \cite{EHT2019DataRelease}. The official EHT data are available from that public release and are not redistributed here. The analysis codes, derived numerical tables, validation records, and figure generation scripts are available from the author upon reasonable request.

\section*{Declaration of competing interest}
The author declares that he has no known competing financial interests or personal relationships that could have appeared to influence the work reported in this paper.

\appendix

\section{Effective-source interpretation}\label{app:source}

For the rotating radial-mass metric, the Einstein tensor was computed independently and projected onto a principal orthonormal frame. The density and principal radial pressure reduce to Eq.~\eqref{eq:rho}. The equality $p_{\parallel}=-\rho$ already rules out a pressureless interpretation. The transverse stresses are angle dependent in the rotating geometry and contain both $m'$ and $m''$. The code was checked against analytic limits and an independent tensor calculation, with a maximum relative discrepancy of about $2.1\times10^{-5}$ over the tested domain.

This effective-source analysis fixes the interpretation of the dark matter branch. It is not a literal rotating cold dark matter solution. Equations~\eqref{eq:nfw}--\eqref{eq:spike} set the enclosed mass scale, and the rotating geometry supplies only a controlled optical response. The parameters stated in the paper reproduce Table~\ref{tab:dmprofiles} and Fig.~\ref{fig:dmcontrol}. The physically motivated corrections are too small to compete with the source and calibration uncertainties in the present EHT analysis.

\section{Closure basis and covariance details}\label{app:closure}

For a scan with baseline vector $\mathbf{v}$, closure quantities are linear combinations of baseline phase and log amplitude in the Gaussian, moderate-signal-to-noise limit. If $\mathbf{A}$ is a full-row-rank closure design matrix, then
\begin{equation}
 \mathbf{y}=\mathbf{A}\mathbf{x},
 \qquad
 \Cmat_y=\mathbf{A}\Cmat_x\mathbf{A}^{\mathsf T}.
\end{equation}
Different independent bases contain the same closure information when their covariance is carried consistently \cite{Blackburn2020}. Redundant closures are not appended as extra independent measurements.

The empirical high/low difference constrains band-differential scatter. It cannot determine a common-mode floor because that term cancels in the difference. The C0/C50/C100 triad is therefore treated as a sensitivity axis rather than a fitted hyperparameter. Every adequacy statement is required to be stable across the triad.

\section{Local projection and boundary calibration}\label{app:projection}

Let $\widetilde{\Jmat}=\Cmat^{-1/2}\partial\muv/\partial\bm{\vartheta}$. A thin singular-value decomposition gives an orthonormal nuisance basis $\mathbf{U}$ and
\begin{equation}
 \Pperp=\mathbf{I}-\mathbf{U}\mathbf{U}^{\mathsf T}.
\end{equation}
The projected finite-$q$ response is Eq.~\eqref{eq:dchi}. Two completed near-Kerr nonlinear nuisance refits at $q/M^2=-0.0005$ and $-0.001$ agreed with the local projection within 0.554\%. This validates the near-Kerr local approximation only. It does not replace a converged full-grid nonlinear profile.

For the one-sided $q\leq0$ branch, the null lies at a boundary. The asymptotic likelihood ratio distribution is the Chernoff mixture $\tfrac12\delta_0+\tfrac12\chi^2_1$ \cite{Chernoff1954}. The corresponding five-percent critical value, $2.705543$, was reproduced by three independent 200,000-draw simulations.

\section{Numerical validation summary}\label{app:numerics}

The production ray tracer uses adaptive Mino-time integration with potential projection. Accepted rays terminate at capture, escape, or after all explicitly retained surface-crossing budgets are complete. The last status is valid only because additional propagation cannot change the truncated surface intensity. All 92 production images have zero unresolved rays.

A radial-step spot check at $N=160$ changed a representative normalized image by total variation $5.03\times10^{-5}$, much smaller than the screen-resolution effect. A padding factor of 8 did not pass the fixed direct-sum Fourier limit. The final factor of 12 is the smallest tested value that did pass, with maximum complex error $9.38\times10^{-4}$.

The convergence failures are summarized in Table~\ref{tab:numerical}. Because the response gate fails independently, no provisional power threshold from the nonconverged libraries is interpreted as an observational or forecast limit.

\Needspace{0.40\textheight}
\section{Permitted and prohibited inference outputs}\label{app:claims}

\refstepcounter{table}
\noindent\begin{minipage}{\textwidth}
\centering
{\small\textsc{TABLE~\thetable.} Scientific interpretation rules for the present frozen analysis.\par}
\vspace{4pt}
{\small
\renewcommand{\arraystretch}{1.13}
\begin{tabular}{@{}l@{\hspace{1.5em}}l@{}}
\toprule
\parbox[t]{0.42\textwidth}{\raggedright Permitted\par} & \parbox[t]{0.42\textwidth}{\raggedright Prohibited\par}\\
\midrule
\parbox[t]{0.42\textwidth}{\raggedright The covariance-aware independent closure likelihood passes the synthetic tests used here.\par} & \parbox[t]{0.42\textwidth}{\raggedright M87* prefers a negative tidal charge.\par}\\[3pt]
\parbox[t]{0.42\textwidth}{\raggedright The frozen Kerr source fails the predeclared absolute adequacy safeguard.\par} & \parbox[t]{0.42\textwidth}{\raggedright The 2017 data constrain $q$ to a numerical interval.\par}\\[3pt]
\parbox[t]{0.42\textwidth}{\raggedright The mismatch is multi-scan and cannot be repaired by one deletion.\par} & \parbox[t]{0.42\textwidth}{\raggedright Any provisional power point from the nonconverged libraries is an EHT or future-array detection floor.\par}\\[3pt]
\parbox[t]{0.42\textwidth}{\raggedright An inadequate source can create a large formal deformation score.\par} & \parbox[t]{0.42\textwidth}{\raggedright This work independently regenerated the official CSV products from UVFITS.\par}\\[3pt]
\parbox[t]{0.42\textwidth}{\raggedright The differential tidal charge response is numerically nonconverged.\par} & \parbox[t]{0.42\textwidth}{\raggedright Fixed screen-plane smoothing is a physical volumetric plasma model.\par}\\[3pt]
\parbox[t]{0.42\textwidth}{\raggedright Absolute image convergence does not imply metric response convergence.\par} & \parbox[t]{0.42\textwidth}{\raggedright The resolution-dependent sign of the deformation score has physical meaning.\par}\\
\bottomrule
\end{tabular}
}
\end{minipage}
\par\vspace{8pt}

\bibliography{refs}

\end{document}